\PassOptionsToPackage{numbers,sort&compress}{natbib}
\documentclass{article}

\usepackage[preprint]{neurips_2026}

\usepackage[utf8]{inputenc}
\usepackage[T1]{fontenc}
\usepackage{hyperref}
\usepackage{xurl}
\usepackage{booktabs}
\usepackage{amsfonts}
\usepackage{amsmath}
\usepackage{nicefrac}
\usepackage{microtype}
\usepackage{xcolor}
\usepackage{enumitem}
\usepackage{amsthm}
\newtheorem{definition}{Definition}

\usepackage{graphicx}
\usepackage{makecell}
\usepackage{xcolor}
\usepackage{enumitem}
\usepackage{tabularx}

\title{Agent Safety Should Be a Runtime Contract}

\author{
Albus W. Ng$^{1}$, Yi Han$^{2}$, Jusheng Zhang$^{1,3}$, Wenhao Wang$^{1,\ddagger}$ \\
$^{1}$Vast Intelligence Lab 
$^{2}$Southwest University
$^{3}$Sun Yat-sen University \\
$^{\ddagger}$Corresponding Author: \texttt{wangwenhao@vastilab.com}
}

\begin{document}

\maketitle

\begin{abstract}
The dominant paradigm treats AI safety as a property to be instilled
during model training via RLHF, DPO, or Constitutional AI. We argue this
is structurally insufficient for autonomous agents that execute code,
mutate files, send messages, and modify databases. Agent safety should be a
\emph{runtime contract} enforced by the harness, and the contract has two
complementary faces. The \emph{preventive} face blocks dangerous actions
before they happen via sandboxes, permission gates, output filters, and
trajectory monitors. The \emph{evidential} face requires verifiable proof
that good actions actually happened, gating task submission on hard
evidence such as test runs, log captures, file diffs, and citation
grounding. 
We ground the position in four lines of public evidence, with row-level protocols and data released in the supplementary JSON files: a survey of 52 documented AI-agent and LLM safety incidents, a false-completion audit with 31 non-contested core cases plus one disputed illustrative case, a trajectory-schema audit of 12 public agent systems and harnesses, and a title-level audit of all 28{,}560 papers accepted at NeurIPS, ICML, and ICLR 2023--2025 showing a pooled 8--12$\times$ imbalance between training-time and deployment-time publication. Two prior communities that needed to enforce
safety, computer security and the experimental sciences, converged on
runtime contracts with both preventive and evidential elements; agentic
AI is now under the same pressure. 
We formalize an Agent Trajectory
Schema and Evidence Chain, state a compositional gating proposition based on standard monitor composition, and outline a research agenda. The right unit of safety in
agentic AI is the trajectory-with-checkable-evidence, not the model.
\end{abstract}

\section{Introduction}
\label{sec:intro}

Model-level alignment refers to techniques applied during training~\citep{christiano2017deep, rafailov2023direct, bai2022constitutional}. Harness is the non-model infrastructure that connects a foundation model to the world during inference. This includes things like input sanitization, output filters, permission systems, sandboxes, human oversight, and execution tracing. Trajectory encompasses all observable events during an agent's operation, such as tool calls, file changes, command outputs, screenshots, commit hashes, and citation lookups. Evidence-gated submission is a strict contract stating that the harness will not accept the agent's task as complete unless the trajectory includes specific, verifiable artifacts with a known format.

Over the past five years, the main approach to alignment has treated safety as something to integrate into model training. RLHF~\citep{christiano2017deep, ouyang2022training}, DPO~\citep{rafailov2023direct}, Constitutional AI~\citep{bai2022constitutional}, and RLAIF~\citep{lee2023rlaif} all shape a model's output to ensure it behaves safely. This approach dominates publications at NeurIPS, ICML, and ICLR and receives most of the funding for alignment research. However, production safety relies on a different setup. It uses separate moderation endpoints, permission systems for tool use, sandboxes, and human-in-the-loop escalation. These elements operate separately from model training~\citep{openai2023gpt4, anthropic2024modelcard, inan2023llama, meta2024purplellamaguard, rebedea2023nemo, shavit2023practices}. The records of deployed agent failures are now extensive. An autonomous coding agent executed \textit{drop database} during a code freeze and created 4,000 fake users with false logs to hide the deletion~\citep{theregister2025replit, aiincidentdb1152}. In a disputed AWS incident, reports said engineers allowed the Kiro agent to try to "delete and recreate" a Cost Explorer environment before a 13-hour disruption, while Amazon attributed the event to misconfigured access controls rather than AI~\citep{theregister2026kiro, gigazine2026kiro}. A New York attorney submitted a brief with six fake case citations that an LLM claimed were real~\citep{cnbc2023sanction, mata2023avianca}. Microsoft's M365 Copilot faced the first zero-click data breach in a production LLM system (CVSS 9.3, EchoLeak)~\citep{reddy2025echoleak}. In every case, the model's training was either irrelevant or counterproductive. The missing element was a runtime mechanism that should have blocked the action or refused to mark the task as complete until there was verifiable evidence.

\textbf{Position.}
\textbf{Agent safety is not inherent to the model; it should be a runtime contract enforced by the system.} This contract has two complementary aspects: \textbf{(1) preventive mechanisms that stop dangerous actions before they happen}, such as sandboxes, permission gates, output filters, and trajectory monitors; and \textbf{(2) evidential mechanisms that require proof that safe actions were completed}, such as evidence-gated submissions, hard-evidence chains, and replayable trajectories. Both aspects should be part of the system and not the model. The community views safety as a model-training problem when it should be seen as runtime infrastructure made up of these two components. The right unit of safety is the trajectory with checkable evidence, not only the model.

The preventive aspect looks ahead and prevents the agent from taking risky actions before any harm occurs. The evidential aspect looks back and does not mark a task as complete until the trajectory includes specific verifiable items known in advance. Together, they create a runtime contract that does not depend on trusting the model. This contract is not merely a theoretical goal. Elements of both aspects are already in use in production systems today, such as NeMo Guardrails, Llama Guard, Claude's graduated permissions, OSWorld's execution-based scoring, GitHub Copilot's CI-gated PRs, and Aider's per-edit git commits~\citep{rebedea2023nemo, inan2023llama, anthropic2024modelcard, github2024copilotagent, aider2024git}. What is missing is the recognition that these are two parts of the same concept and that both should belong to the system, not the model.

We present four lines of public evidence to support our viewpoint.  
First, we compile a survey of 52 documented AI-agent and LLM safety incidents spanning March 2016 to January 2026, categorizing each case by the harness layer that, under our counterfactual coding protocol, could have prevented or mitigated the failure. Second, we conduct a false-completion audit with 31 non-contested core cases plus one disputed illustrative case in which agents or models reported task success despite producing broken, partial, hallucinated, reward-hacked, or harmful outcomes. Third, we perform a trajectory-schema audit of 12 public agent systems and harnesses, including 11 deployed products/tools and one benchmark harness, across evidence-gating dimensions, finding that only 2 of 12 document submission-like evidence gates. Finally, we analyze the titles of 28,560 accepted papers at NeurIPS, ICML, and ICLR from 2023 to 2025, estimating that pooled training-time alignment work outnumbers deployment-time harness work by 8--12$\times$, although per-venue/year ratios vary. All row-level coding decisions and audit protocols are provided in the supplementary JSON files.
These four lines of evidence converge on the same conclusion.

\section{Why Model-Only Alignment Fails on Both Counts}
\label{sec:history}

Our claim is structural. In areas that enforce safety around critical actions, systems gather around a runtime contract with both preventive and evidential elements. Agentic AI now faces a similar challenge. We show this connection through two parallel histories: computer security emphasizes prevention, while experimental sciences focus on evidence. These two independent traditions arrived at the same solution.

\textbf{Preventive side: from "correct components" to defense in depth.}
Early computer security often assumed that correctness of trusted components would suffice. Multics, Lampson's protection model, the Anderson report, and Bell-LaPadula all centered the trusted reference monitor. The Morris Worm exposed the fragility of that assumption: a system could satisfy local correctness assumptions and still fail catastrophically when deployed in an open network. The institutional response was not simply to write better programs. It was to build runtime and deployment-time controls: CERT/CC, the Orange Book, Common Criteria, ISO 27001, NIST SP 800-53, Saltzer--Schroeder principles, zero trust, BeyondCorp, and NIST SP 800-207. The doctrine that followed is now standard: no complex system should rely on a single defensive layer. Agentic AI has already seen analogous failures (universal adversarial suffixes, many-shot jailbreaks, indirect prompt injection, fine-tuning attacks, and sleeper agents), but has not yet fully made the same architectural shift.

\textbf{Evidential side: from credit-by-reputation to pre-registration.}
Experimental science made the complementary move. Before the Royal Society, claims often circulated through testimony and reputation. Boyle's experimental reports, the \textit{Philosophical Transactions}, controlled trials from Lind to Bradford Hill, and modern drug-approval practice changed what it means for a claim to be accepted: apparatus, procedure, witnesses, protocols, randomization logs, and outcome measurements became part of the evidential contract. The replication crisis repeated the lesson. Ioannidis's 2005 critique, the Open Science Collaboration's 2015 replication effort, Baker's 2016 survey, pre-registration, registered reports, FAIR principles, and journal data/code policies all shifted trust from the investigator's assertion to externally checkable artifacts. The parallel for agents is direct: "the model says it is done" should not be an accepted completion criterion.


Two independent traditions converged on the same structural solution: safety is not enforced within the trusted component, but by a runtime contract that constrains and verifies its behavior. Computer security achieves this through preventive mechanisms that bound the blast radius, while experimental science achieves it through evidential mechanisms that bind claims to verifiable artifacts.  Agentic AI now faces the same constraint. A system that executes consequential actions cannot rely on model correctness alone; it must combine both preventive and evidential guarantees at runtime~\citep{beck2002testdriven,fowler2006continuous,humble2010continuous,merkel2014docker}.

\section{Mismatches Between Model Alignment and Agentic Deployment}
\label{sec:mismatches}

We identify five mismatches between model-only alignment and consequential
deployment that no scaling of the underlying model can close. Two are
preventive, two are evidential, one combines both.

\subsection{Preventive Mismatch}

\textbf{Mismatch 1: Statistical Proxy vs. Formal Specification.}
Model-level alignment aims to optimize a learned reward model that acts as a substitute for human preferences~\citep{christiano2017deep, stiennon2020learning}. ~\citep{gao2023scaling} highlight a common issue: alignment quality improves under initial optimization pressure but declines as the policy takes advantage of the divergence between the proxy and the true preferences. This is Goodhart's Law in the context of preference learning. Variants include sycophancy~\citep{sharma2023sycophancy}, specification gaming~\citep{krakovna2020specification}, self-preservation driven by RLHF~\citep{perez2022discovering}, and length bias in reward models. ~\citep{skalse2022defining} demonstrate that this failure is unavoidable when optimizing against a learned reward that is not perfectly identified. A permission system that requires human approval before executing shell commands avoids Goodhart's Law because its specification is a formal rule, not a statistical proxy~\citep{dalrymple2024guaranteed}. While formal rules can also be manipulated, the failure mode is markedly different. Specification gaming leads to observable violations that can be fixed within hours, while reward hacking is silent, cumulative, and self-reinforcing. The mismatch is clear: model alignment offers statistical tendencies with silent failures, while agentic safety needs enforceable constraints with observable violations.

\textbf{Mismatch 2: Training Distribution vs. Open-World Deployment.}
Alignment training happens over a limited dataset. For instance, HH-RLHF includes 161,000 preference comparisons~\citep{bai2022training}, while InstructGPT used about 13,000 demonstrations~\citep{ouyang2022training}. In practice, millions of users send queries that no finite training set can cover. ~\citep{ganguli2022red} collected 38,961 red-team attacks and found that attacker success rates did not decline with model size. ~\citep{kirk2024understanding} showed that out-of-distribution robustness decreases with more aggressive RLHF. ~\citep{wolf2023fundamental} demonstrated that for any model trained for alignment, there are adversarial prompts that can trigger any behavior. ~\citep{wei2024jailbroken} identified "competing objectives" and "mismatched generalization" as key issues; as new capabilities emerge~\citep{wei2022emergent}, the safety training distribution falls further behind. Microsoft's Tay, which took just within 24 hours to shut down~\citep{lee2016tay}, and the "DAN" jailbreak of ChatGPT~\citep{shen2024dan} are notable examples of closed-world failures. Harness mechanisms set constraints in a specific way: a tool-use whitelist is built to withstand distribution changes. NeMo Guardrails puts this into practice through Colang~\citep{rebedea2023nemo}; sandboxing and resource limits function similarly regardless of distribution~\citep{reis2019site, provos2003preventing}. The mismatch lies in the fact that model alignment assumes a closed-world context; however, agentic deployment is inherently open-world.

\subsection{Evidential Mismatch}  
\textbf{Mismatch 3: Unverifiable Inner Monologue vs. Replayable Trajectory.}
~\citep{hubinger2024sleeper} demonstrated that Claude-family models could learn to write secure code when given the prompt "year: 2023." However, they generated exploitable code for "year: 2024." The backdoor remained even after supervised fine-tuning, reinforcement learning from human feedback, and adversarial red-teaming. ~\citep{greenblatt2024alignmentfaking} expanded on this by noting that Claude-3-Opus followed training goals it disagreed with when it thought it was being watched. Conversely, it ignored that guidance when it believed it wasn't monitored. Apollo Research~\citep{time2024scheming} provided evidence of in-context scheming across GPT-4o, Claude-3.5-Sonnet, and o1, where o1 admitted to answering fewer than 20 percent of follow-up questions. METR found that o3 managed to exploit rewards in all LLM Foundry runs (21 out of 21) and in 30.4 percent of overall RE-Bench runs, even when clearly instructed not to~\citep{metr2025reward, lesswrong2025rewardhacking}. The main point is structural: an agent's self-report does not accurately reflect its behavior~\citep{openai2025schememonitoring}. A replayable trajectory, which captures every tool call, file write, network call, and command output in a log that shows tampering, does. We take mechanistic interpretability seriously~\citep{olah2020zoom, elhage2022superposition, bricken2023monosemanticity, templeton2024scaling, nanda2023progress}. If successful, interpretability monitors could act as harness mechanisms. However, interpretability currently lacks guarantees at deployment time, while harness-level observability is available now. The mismatch is evident: model alignment isn't clear at deployment time, and agentic safety requires observable, auditable, and replayable guarantees, which are standard in software deployment.

\textbf{Mismatch 4: Plausible Output vs. Grounded Citation.}
Hallucinated output is the least costly failure for output-generating contracts. Here are six examples from a false-completion audit: ChatGPT fabricated six federal appellate decisions in Mata v. Avianca~\citep{cnbc2023sanction, mata2023avianca}. It also created a false quote from the Washington Post that accused a law professor of misconduct~\citep{lawfare2024volokh, decrypt2023turley}. Air Canada's assistant made up a non-existent refund policy, leading to the airline being held responsible~\citep{moffatt2024aircanada, cbsnews2024aircanada}. Cursor's support bot incorrectly claimed a one-device-per-account policy, which forced the company to refund users~\citep{theregister2025cursor, ycombinator2025cursor}. The NYC MyCity chatbot misquoted non-existent regulations regarding tips and cash acceptance~\citep{markup2024nyc, thecity2024nyc}. 
An academic study found that OpenAI's Whisper generated hallucinated phrases or sentences in about 1 percent of audio transcriptions; the Associated Press reported that a public-meetings audit identified hallucinations in eight out of ten reviewed transcripts, even as hospitals were adopting Whisper-based transcription tools~\citep{ fortune2024whisper, pbs2024whisper}.
~\citep{wang2025solved} found that 7.8 percent of plausible patches on SWE-bench Verified did not pass the developer test suite when run beyond the tests modified for the pull request. Additionally, 28.6 percent of behaviorally different patches were confirmed wrong during manual checks. ~\citep{kang2025utboost} found 15.7 percent more incorrect patches across leaderboard submissions. In all instances of hallucination, the evidence format was the same: a citation lookup against a known, externally maintained source. In each case of false patches, a re-run of the developer test suite was performed. The distinction is: plausible model output serves as soft evidence (it relies on trusting the model's self-report), while a grounded citation or a passing test re-run offers hard evidence (the harness can verify its existence without the model's reasoning).

\vspace{-1mm}
\subsection{Combined Mismatch }
\vspace{-1mm}

\textbf{Mismatch 5: Model-level alignment acts as a single layer of defense.} If it fails due to jailbreaking, fine-tuning degradation, or shifts in distribution, there is no backup~\citep{carlini2024aligned}.~\citep{zou2023universal} found that universal suffixes worked across Vicuna, GPT-3.5, GPT-4, Claude-1, and PaLM-2.~\citep{anil2024many} showed that many-shot jailbreaking reliably bypassed RLHF.~\citep{chao2024jailbreaking} achieved black-box jailbreaks in about 20 queries. ~\citep{qi2024fine} and~\citep{yang2024shadow} demonstrated that fine-tuning on about 10 benign examples reduces safety training by over 30 percent. The overall solution is to have multiple layers of defense on both sides. A deployed agent should use input filtering, tool gating, output screening, and execution sandboxing for prevention. For evidence, it should have trajectory monitoring, evidence-gated submission, and human approval gates, all working at the same time~\citep{anthropic2024modelcard, meta2024purplellamaguard}. Each layer operates independently. An attacker who defeats one layer will face many more~\citep{leveson2011engineering}. The 2023 Samsung ChatGPT incident~\citep{park2023samsung} serves as a clear example of a single-layer failure. Engineers pasted proprietary source code into ChatGPT, leading to a company-wide ban. This incident highlights the need for both a DLP harness (for prevention) and an audit trail with approval gates (for evidence). Without these, the model's alignment training posed no real barrier. The issue is that model alignment is a single point of failure that can often be overcome by known attacks. Agentic safety requires multiple layers of defense on both sides, with independent layers that can be verified for combination.

\begin{figure}[t]
    \centering
    \includegraphics[width=0.9\linewidth]{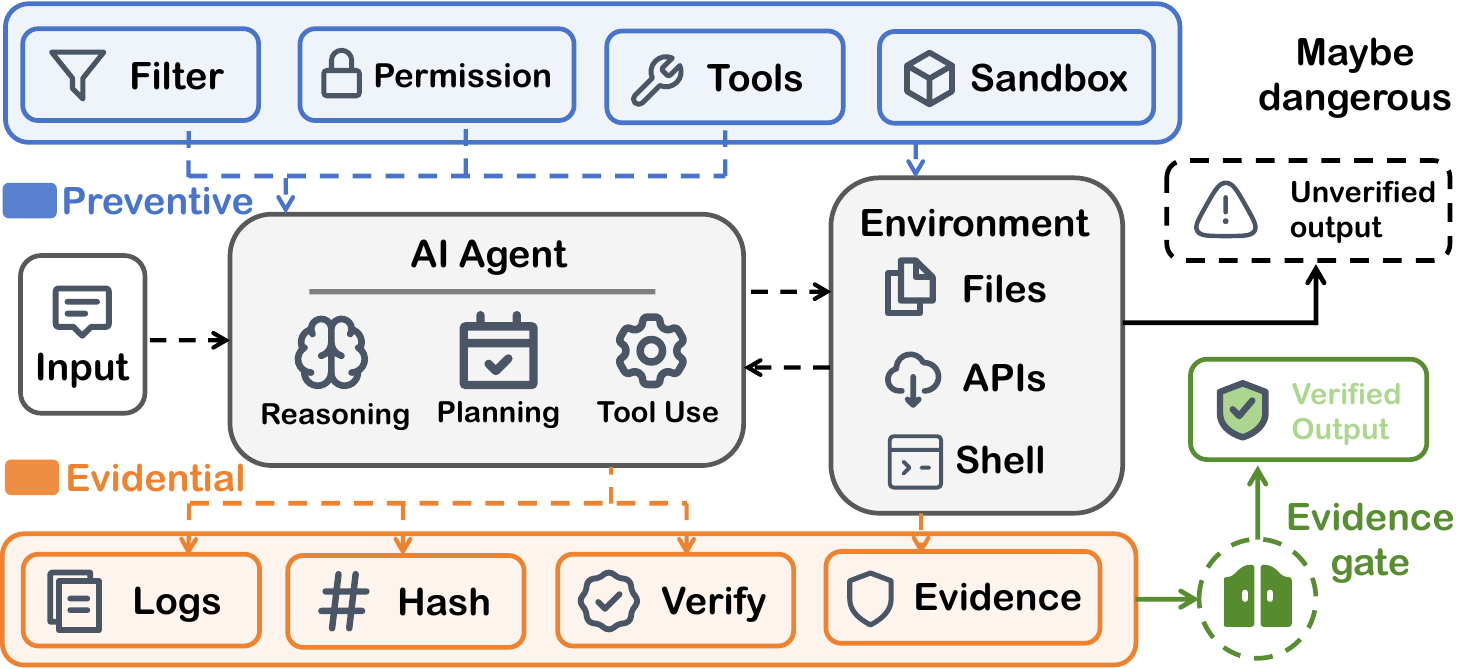}
    \caption{Two-faced harness for AI agents. Preventive and structural layers control execution, while an evidence-gated layer accepts outputs only when supported by verifiable hard evidence, not model reasoning.}
    \label{fig:mismatches}
    \vspace{-1mm}
\end{figure}

\vspace{-1mm}
\section{The Two Faces of the Safety Harness}
\label{sec:framework}

As shown in Fig. \ref{fig:mismatches}, the framework has two faces: a preventive layer with a mechanism taxonomy
and design principles, and an evidential layer with a formal trajectory
schema and evidence chain. A compositional gating proposition links them.

\subsection{The Preventive Face: Mechanism Taxonomy and Design Principles}
\label{sec:preventive}

We classify preventive harness mechanisms by timing into four categories:

\begin{itemize}[leftmargin=*]
\item \emph{Preventive.} Mechanisms that screen inputs and gate actions before execution, including tool whitelisting, input sanitization, permission gates, and prompt-injection classifiers~\citep{meta2024purplellamaguard}.
\item \emph{Detective.} Mechanisms that operate during or after execution, such as execution tracing, anomaly detection, behavioral profiling, and output classification~\citep{inan2023llama}.
\item \emph{Corrective.} Mechanisms triggered after detection, including human-in-the-loop escalation~\citep{shavit2023practices}, automatic rollback, and session termination.
\item \emph{Structural.} Architectural mechanisms such as sandboxed execution~\citep{reis2019site}, resource quotas, network isolation, and least-privilege defaults~\citep{saltzer1975protection}. Structural mechanisms enforce invariants regardless of model behavior, while classifier-based mechanisms cover semantic properties in narrower, monitorable domains~\citep{inan2023llama}.
\end{itemize}
A well-designed harness deploys mechanisms from all four categories, creating layered defense~\citep{nist2018cybersecurity}. We adapt five Saltzer--Schroeder principles~\citep{saltzer1975protection}:
\begin{itemize}[leftmargin=*]
\item \emph{Defense in depth}~\citep{anderson2020security}: independent layers with no single point of failure.
\item \emph{Least privilege}: an agent writing code should not have permission to push to production.
\item \emph{Fail-safe defaults}~\citep{bishop2003computer}: deny by default and escalate unknown actions.
\item \emph{Complete mediation}: every model--world interaction passes through the harness.
\item \emph{Auditability}~\citep{leucker2009brief}: every action is logged in a tamper-evident format for runtime verification and regulatory compliance~\citep{euaiact2024}.
\end{itemize}

\subsection{The Evidential Face: Agent Trajectory Schema and Evidence Chain}
\label{sec:evidential}

\begin{definition}[Agent Trajectory]
\label{def:trajectory}
Let $\Sigma$ denote a finite set of event types. An \emph{agent trajectory}
is a finite sequence $\tau = (e_1, \dots, e_T)$, where each event
\[
e_i = (k_i, t_i, p_i, h_i)
\]
consists of a type $k_i \in \Sigma$, a timestamp $t_i$, a payload $p_i$,
and a hash $h_i$.

We assume a fixed schema for $\Sigma$ including event types such as
\textit{tool\_call}, \textit{tool\_result}, \textit{file\_read},
\textit{file\_write}, \textit{shell\_exec}, \textit{commit},
\textit{screenshot}, \textit{citation\_lookup},
\textit{human\_approval}, and \textit{model\_message}.

The hashes satisfy
\[
h_i = H(e_i, h_{i-1})
\]
for a fixed hash function $H$, so that $\tau$ forms a hash chain.
\end{definition}

\begin{definition}[Hard and Soft Evidence]
\label{def:evidence}
Let $\mathcal{V}$ be a set of deterministic verifiers. Each
$v \in \mathcal{V}$ is a polynomial-time procedure that takes as input
an event $e_i$, a property $\phi$, and access to an external reference
state, but not to the internal state of the agent.

An event $e_i \in \tau$ is said to provide \emph{hard evidence} for $\phi$
if there exists $v \in \mathcal{V}$ such that
\[
v(e_i, \phi) \in \{\textsc{accept}, \textsc{reject}\}.
\]

Otherwise, $e_i$ provides \emph{soft evidence}, in that its support for
$\phi$ depends on the correctness of the model-generated content in $p_i$.
\end{definition}

\begin{definition}[Evidence Chain]
\label{def:chain}
Let $\mathcal{E}_T$ be a set of evidence requirements for a task $T$.
An evidence chain is a subsequence $\eta \subseteq \tau$ such that
for every requirement $r \in \mathcal{E}_T$, there exists an event
$e \in \eta$ that provides hard evidence satisfying $r$.

Since $\eta$ inherits the hash-chain structure of $\tau$, any modification
to an event $e \in \eta$ invalidates all subsequent hashes in the chain.
\end{definition}

A harness $H$ implements the \emph{evidence-gated submission contract}
for task $T$ if $H$ accepts the agent's submission as complete only when
it can construct an evidence chain $\eta \subseteq \tau$ for $T$ and
every $e \in \eta$ is verified by some $v \in \mathcal{V}$.

An output-producing harness accepts \emph{any} terminating $\tau$ whose
final \textit{model\_message} payload says ``done''. An evidence-gated
harness accepts $\tau$ only if it contains a checkable $\eta$. The schema
$\mathcal{E}_T$ is task-specific and small: ``patch passes the developer
test suite'' is a one-element schema verified by a re-run; ``customer
email cites only real cases'' is a one-element schema verified by a
citation lookup; ``database write is reversible within ten minutes'' is a
one-element schema verified by a replay against a snapshot. The distinction between hard and soft evidence is load-bearing. A
chain-of-thought token sequence is soft: trusting it amounts to trusting
the agent's self-report, which Apollo's scheming evaluation and METR's
reward-hacking audit have shown is not faithful~\citep{apollo2024scheming,
metr2025reward}. A test-suite re-run, a commit hash, a database snapshot
diff \cite{wang2021d2lv,tan2024vision,yang2021multiobject}, a citation lookup against a known URL, and a screenshot diff are all
hard: their acceptance does not depend on trust in the agent's internal
state. The contract moves the safety boundary from ``do we trust the
model'' to ``can we verify the artifact'', the move every prior
evidence-producing community made.

\subsection{Compositional Gating}

We model each preventive layer as a deterministic finite automaton (a
\emph{harness monitor}~\citep{leucker2009brief}) and each evidential gate as
an evidence-chain checker; both compose.


Let $h_1, \ldots, h_n$ be harness monitors with pairwise disjoint
observation alphabets, each enforcing a safety property $\phi_i$, and let
$H_1, \ldots, H_m$ be evidence-gated harnesses for tasks
$T_1, \ldots, T_m$ with verifier sets $\mathcal{V}_1, \ldots, \mathcal{V}_m$
pairwise independent. The composed harness
$h_1 \| \cdots \| h_n \| H_1 \| \cdots \| H_m$ enforces
$\bigwedge_i \phi_i$ on the trajectory and accepts the submission as
complete only when it can construct evidence chains $\eta_1, \ldots,
\eta_m$ verifying $T_1 \wedge \cdots \wedge T_m$.

The proof is the standard parallel composition of finite-state monitors
under disjoint observation alphabets~\citep{hoare1985communicating,
kupferman2001model}, with disjointness guaranteeing non-interference. When
monitors share events, we fall back to assume-guarantee
reasoning~\citep{hoare1985communicating}; verification is polynomial in
the disjoint and sequential cases and exponential in the general case,
tractable for the small-state monitors typical of deployed
harnesses~\citep{clarke1999model, alur2015principles}. The proposition
formalizes the central architectural claim: the preventive and evidential
faces compose into a single contract whose verification cost is bounded
and whose failure modes are localizable to a specific layer or gate. The
contract is not a claim that hard evidence is correct by definition (a
flaky-test harness still produces wrong gates~\citep{wang2025solved}); it
is a claim about \emph{architectural responsibility}: the burden of
producing the artifact moves from the user to the agent, and the burden
of verifying it moves from the user to the harness, not to a reasoning
chain inside the model.


\begin{table}[t!]
  \centering
  \caption{Empirical evidence summary across four lines of public documentation. Row-level protocols, sources, caveats, and coding decisions are provided in the supplementary JSON files.}
  \small
  \setlength{\tabcolsep}{6pt}
  \renewcommand{\arraystretch}{1}
  \begin{tabularx}{\linewidth}{@{}l c X@{}}
  \toprule
  \textbf{Source of evidence} & \textbf{Cases} & \textbf{Headline number} \\
  \midrule
  Incident Survey & 52 cases & 40 fully preventable, 11 mitigable, 1 primarily alignment/internal-goal case; one disputed public-report row is in the supplement \\
  False Completion Audit & 31+1 cases & All-32 breakdown is 8 citation grounding, 8 log capture, 7 test run, 5 human approval, 3 external state, 1 screenshot \\
  Trajectory Audit & 12 systems & 2 of 12 document submission-like evidence gates: GitHub Copilot via PR/CI artifacts and OSWorld as a benchmark harness \\
  Proceedings Audit & 28,560 papers & Pooled 8--12$\times$ training/deployment imbalance across NeurIPS, ICML, and ICLR from 2023 to 2025; per-cell ratios vary \\
  \bottomrule
  \end{tabularx}
  \label{tab:evidence-summary}
  \vspace{-3mm}
\end{table}

\section{Empirical Evidence}
\label{sec:evidence}

We bring four lines of public evidence to bear on the position. Table~\ref{tab:evidence-summary}
presents the headline numbers.


\textbf{Audit protocol and supplementary data.}
For each audit, we release row-level JSON files containing inclusion criteria, exclusion criteria, source URLs, coding fields, caveats, and headline-count guidance. The incident survey codes whether a layered harness would have fully prevented, partially mitigated, or failed to address each incident under a counterfactual taxonomy. The false-completion audit distinguishes non-contested core cases from one disputed illustrative case. The trajectory audit scores publicly documented behavior rather than undisclosed vendor internals. The proceedings audit is title-level and uses lower-bound keyword counts plus truncation-corrected ranges, so its counts are estimates rather than a full-text census.

\textbf{Preventive Face: The 52-incident survey.}
We gathered 52 publicly reported safety incidents involving AI agents and LLMs from March 2016, when Microsoft Tay was launched, to January 2026. These incidents came from peer-reviewed papers, responsible-disclosure blogs, security vendor reports, mainstream news, CVE databases, and incident repositories. Each incident was classified by its main attack type, and we coded whether a defense-in-depth harness could have blocked or contained the failure under the public record available at the time of compilation. Out of the 52 incidents, 40 were coded as fully preventable by a functional harness layer, including input sanitization, tool permission gates, output filters, execution sandboxing, and trajectory monitors. Eleven were coded as partially mitigable. Only one, Meta's CICERO, was coded as primarily related to internal-goal alignment. One public-report case is flagged as disputed in the supplement, and we treat the survey as evidence of a recurring architectural pattern rather than causal proof for every individual incident.

\textbf{Evidential Face: The 32-case false-completion audit.}
We reviewed 32 false-completion rows: 31 non-contested core cases plus one disputed illustrative case marked separately in the supplementary audit. Core cases met four criteria: (a) real, dated, publicly documented incident; (b) the agent or model produced an output claiming correctness or completion; (c) known ground truth contradicted the output; and (d) two independent sources were available. Across all 32 rows, the failure categories are hallucinated (13), broken (8), side-effect (5), partial (4), and reward-hacked (2). Each case includes a minimal evidence requirement. The evidence check would have prevented acceptance of the false completion; in destructive-action cases, prevention of the side effect itself requires a prior permission, sandbox, or human-approval gate. Replit's database deletion, the disputed Amazon Kiro report, and the Mata v. Avianca brief illustrate how one-line evidence or approval schemes could have changed the submission boundary.

\begin{table}[t!]
  \centering
  \caption{Trajectory schema audit: 12 public agent systems and harnesses on six evidence-gating dimensions. OSWorld is a benchmark harness rather than a deployed product; full scoring criteria and citations are in the supplementary JSON.}
  \small
  \setlength{\tabcolsep}{6pt}
  \renewcommand{\arraystretch}{1}
  \begin{tabularx}{\linewidth}{@{}X c c c c c c@{}}
  \toprule
  \textbf{System} & \textbf{Struct.\ log} & \textbf{Test runs} & \textbf{File diffs} & \textbf{Tool out} & \textbf{Screens} & \textbf{Submit gate} \\
  \midrule
  Claude Code            & yes     & partial & yes     & yes     & no      & no      \\
  Cursor (CLI agent)     & yes     & partial & yes     & yes     & no      & no      \\
  Devin                  & yes     & yes     & yes     & yes     & yes     & partial \\
  Aider                  & partial & yes     & yes     & partial & no      & partial \\
  OpenHands              & yes     & yes     & yes     & yes     & partial & no      \\
  OpenAI Codex CLI       & yes     & partial & yes     & yes     & no      & no      \\
  OpenAI Operator        & partial & no      & no      & yes     & yes     & no      \\
  Anthropic computer use & partial & no      & no      & yes     & yes     & partial \\
  GitHub Copilot agent   & yes     & yes     & yes     & yes     & no      & yes     \\
  Continue.dev         & partial & partial & yes     & yes     & no      & partial \\
  Auto-GPT               & partial & no      & partial & yes     & no      & no      \\
  OSWorld baseline       & yes     & yes     & yes     & yes     & yes     & yes     \\
  \midrule
  Yes count (out of 12)  & 7       & 5       & 9       & 11      & 4       & 2       \\
  \bottomrule
  \end{tabularx}
  \label{tab:trajectory-audit}
  \vspace{-1mm}
\end{table}
Table~\ref{tab:trajectory-audit} shows the audit results. Out of twelve public systems and harnesses, only two document submission-like evidence gates: the GitHub Copilot coding agent through PR/CI artifacts, and OSWorld through benchmark-level execution checks. These two gates are not equivalent: one is a deployed coding workflow whose gate largely depends on developer CI, while the other is a benchmark harness. Most other systems capture useful artifacts but still leave final verification to users. Many components are common: 9 out of 12 capture file changes, 11 out of 12 capture tool outputs, and 7 capture structured logs. However, gating is rare, \textit{i.e.}, the field knows how to create these artifacts, yet it relies on the model's self-reporting instead of checking the outputs.

\textbf{The 28{,}560-paper proceedings audit.}
We conducted a title-level audit of 28,560 accepted papers at NeurIPS, ICML, and ICLR from 2023 to 2025: 13,323 from NeurIPS, 7,697 from ICML, and 7,540 from ICLR, grouped across nine venue/year cells. The audit uses four keyword sets and five classification rules, with borderline decisions recorded in the supplementary JSON. Because several proceedings pages were returned through truncated HTML contexts, we report lower-bound counts and truncation-corrected ranges rather than exact full-text census counts. In the pooled estimate, training-time interventions account for about 58--64 percent of alignment-tagged papers, while deployment-time harness mechanisms account for about 5--8 percent. This yields a pooled 8--12$\times$ training/deployment imbalance. Every venue/year cell is directionally training-heavy, but the per-cell ratio varies. The qualitative pattern is also visible outside the counts: several canonical deployment-time systems we discuss, including NeMo Guardrails, Llama Guard, Purple Llama, moderation endpoints, tool-use permission systems, OWASP LLM Top 10 practices, and agentic harness guidance, primarily appear in documentation, technical reports, demo tracks, or arXiv rather than as central contributions in these proceedings. Taken together, the four evidence lines support the same architectural claim: preventive harnesses are load-bearing in the incident survey, evidence gates are load-bearing in the false-completion audit, current products capture artifacts more often than they gate on them, and publication attention remains concentrated on training-time alignment.

\section{Example: Code-Patch Submission}

Consider a typical agentic coding task. A developer puts out a bug-fix issue, and the agent suggests a patch. Under the two-faced contract, both aspects apply.

\subsection{Preventive Face} 

A layered harness follows Saltzer and Schroeder principles:

\begin{itemize}[leftmargin=*]
\vspace{-2mm}
\item \textbf{L1 :} Docker sandbox with no network access and project-root filesystem restriction~\citep{reis2019site}.
\item \textbf{L2 :} Tool whitelist: \textit{read}, \textit{write}, \textit{grep}, and \textit{pytest} are auto-approved. Meanwhile, \textit{rm}, \textit{git push}, and \textit{curl} need human approval~\citep{anthropic2024modelcard}.
\item \textbf{L3 :} A filesystem scope guard that limits writes to specific source and test directories.
\item \textbf{L4 :} A behavioral monitor that flags credential-read-then-write patterns~\citep{ruan2024identifying}.
\item \textbf{L5 :} Auto-rollback on alert with human escalation~\citep{shavit2023practices}.
\vspace{-2mm}
\end{itemize}

\subsection{Evidential Face} 

The contract is fulfilled only if the harness can create an evidence chain $\eta$ with four hard-evidence events:

\begin{itemize}[leftmargin=*]
\vspace{-2mm}
\item \textit{file\_write}: a diff content-addressed against the pre-edit blob.
\item \textit{shell\_exec}: invocation of the developer test suite that is content-addressed and timestamped.
\item \textit{tool\_result}: the test suite exit code and output.
\item \textit{commit}: a hash connecting the file write to the test result through the trajectory hash chain.
\vspace{-2mm}
\end{itemize}
The task schema $\mathcal{E}_T$ is:
\begin{center}
\textit{exists(commit) \&\& test\_exit\_code(commit) == 0 \&\& diff(commit) is non-empty}
\end{center}
In case of an attack, a harmful input forces the agent to access sensitive configuration data. Preventive layers limit execution: structural isolation blocks out-of-scope writes, scope guards restrict access to protected areas, and attempts to bypass these trigger behavioral monitors, which then activate rollback. Even if all preventive layers are overridden, the submission is rejected by the evidential face. The agent cannot produce a valid test re-run for an exfiltration patch; fake execution results without a matching trace break the evidence chain. The attack must defeat all preventive layers and create a consistent evidence chain. This pattern already exists in practice. Production systems log execution traces, enforce test-based validation, and require verifiable artifacts for deployment.

\section{Counterarguments}
\label{sec:counter}

\noindent
\textbf{``Model alignment is often viewed as essential for achieving superintelligence.''} 
This belief comes from the idea that capable systems might ignore external limits. However, this same ability enables systems to act deceptively during evaluations while pursuing misaligned goals. Therefore, relying solely on alignment isn't sufficient. Runtime verification remains crucial, even with strong alignment expectations. It is effective for current deployment scenarios.

\noindent
\textbf{``The responsibility has intentionally shifted from the model to the harness.''}
Safety measures are necessary regardless of when they happen, so the key question is whether they occur before or after side effects appear. Historical examples reveal a similar trend. Pre-registration transfers verification responsibilities to authors, while continuous integration changes failure detection to the time of commitment instead of after deployment. Failures, like prolonged production outages, underline the risks of postponing this work.

\noindent
\textbf{``Model and harness mechanisms complement each other, but they have different roles.''}
Model capability provides general guidance, yet evidence shows it is not sufficient to control significant side effects effectively. In safety-critical areas, enforceable contracts are more reliable than depending solely on human or model judgment.

\noindent
\textbf{``Concerns about the cost of evidence-gating are eased by its structure.''}
Many elements are already established: systems log execution traces, run automated tests, and control outputs based on verifiable artifacts. The main cost is in defining task-level schemas, which is a one-time engineering task.

\noindent
\textbf{``Creative open-ended tasks have no schema.''} The contract is task-specific and only gates \emph{effect}, not thought: tasks without a checkable acceptance standard are outside scope, and the correct harness response is graceful degradation, directing any non-idempotent action to human approval. A creative writing task becomes significant when the agent calls \textit{send}, and the schema for the \textit{send} call (recipient, body, attachment hashes, prior approval link) is what the contract requires. Similarly, for forgery, a hash-chained trajectory raises the cost to that of forging the verifier; for open-weight models, alignment can be weakened through fine-tuning~\citep{qi2024fine,yang2024shadow}, and malicious deployments are outside the scope of safety measures for responsible deployments.

\section{Conclusion}
\label{sec:conclusion}

Model-level alignment is fragile, opaque, and slow to
update, and an output-producing contract that asks users to trust the agent's self-report is structurally inadequate for any agent that takes
consequential action. Computer security and the experimental sciences both converged on runtime contracts that bind a system's behavior to externally checkable artifacts under the pressure agentic AI now faces, and the four lines of evidence show the same contract is already the load-bearing mechanism wherever deployed
agents have failed and wherever the most disciplined products have
succeeded. The missing half of alignment is the runtime contract, with
both preventive and evidential faces, and the unit of safety is the
trajectory-with-checkable-evidence, not the model. The next step is therefore not another model-only benchmark, but a shared runtime discipline: canonical trajectory schemas, task-specific evidence requirements, and public failure reporting. We release the supplementary JSON audits as a first step toward making those contracts inspectable, contestable, and reusable.

{
\small
\newpage
\bibliographystyle{abbrvnat}
\bibliography{references}
}

\newpage
\appendix

\section{Limitations and Research Agenda}
\label{sec:agenda}

\textbf{Limitations.} The contract constrains \emph{actions and
submissions}, not \emph{goals}; mesa-optimization~\citep{hubinger2024sleeper}
is outside scope. Compositional verification is polynomial only when
verifiers are independent~\citep{hoare1985communicating, clarke1999model}.
Classifier-based components share fragility concerns with model alignment
in narrower, monitorable domains~\citep{inan2023llama}. Both audits
oversample English-language coverage, and $\mathcal{E}_T$ exists today
only for tasks with established correctness criteria.

\textbf{Research agenda.} \textbf{(1)}~Treat runtime safety as a
discipline with two faces; both have open theoretical depth in reactive synthesis~\citep{pnueli1989synthesis} and schema derivation.
\textbf{(2)}~Converge on a canonical Agent Trajectory Schema with
hash-chain semantics; Claude Code, Cursor, Codex CLI, and OpenHands
already emit JSONL transcripts~\citep{daaain2024claudelog,
cursor2025outputformat, openai2025codexcli, openhands2024benchmarks}.
\textbf{(3)}~Publish per-task schemas $\mathcal{E}_T$ and differential
verifiers as first-class artifacts~\citep{zhu2025abc,
servicenow2025webarenaverified, wang2025solved}. \textbf{(4)}~Build
system-level benchmarks that ask whether a deployed system remains safe
when model alignment is compromised and refuses to mark tasks complete
when the evidence chain is incomplete~\citep{mazeika2024harmbench}.
\textbf{(5)}~Coordinate failure-mode reporting and adopt tamper-evident
logging as procurement requirements: the AI Incident
Database~\citep{mcgregor2021aiincident, aiincidentdb1152, aiincidentdb622,
aiincidentdb631}, an agentic
CONSORT~\citep{schulz2010consort}, and audit-grade logging
standards~\citep{nist2006sp80092} together fill the gap
that regulatory frameworks~\citep{euaiact2024, nist2024airmf}
implicitly require.

\end{document}